\documentclass[10pt]{iopart}
\usepackage{graphicx}
\usepackage{latexsym}
\usepackage{dcolumn}
\usepackage{bm}
\usepackage{amssymb}
\usepackage{cite}
\usepackage{hyperref}

\newcommand{\be}{\begin{equation}}
\newcommand{\ee}{\end{equation}}
\newcommand{\bea}{\begin{eqnarray}}
\newcommand{\eea}{\end{eqnarray}}
\newcommand{\bse}{\begin{subequations}}
\newcommand{\ese}{\end{subequations}}

\begin{document}

\paper[]{Dynamics of fluctuations in the Gaussian model with conserved dynamics}
\date{\today}

\author{Federico Corberi$^{1,2}$, Onofrio Mazzarisi$^{1}$,\\ Andrea Gambassi$^{3,4}$}

\address{$^{1}$ Dipartimento di Fisica ``E.~R. Caianiello'', Universit\`a  di Salerno, via Giovanni Paolo II 132, 84084 Fisciano (SA), Italy.}
\address{$^{2}$ INFN, Gruppo Collegato di Salerno, 
and CNISM, Unit\`a di Salerno, Universit\`a  di Salerno, via Giovanni Paolo II 132, 84084 Fisciano (SA), Italy.}
\address{$^{3}$ SISSA - International School for Advanced Studies, via Bonomea 265, 34136 Trieste, Italy.}
\address{$^{4}$ INFN, Sezione di Trieste, via Bonomea 265, 34136, Trieste, Italy.}

\ead{corberi@sa.infn.it, omazzarisi@unisa.it, gambassi@sissa.it}

\begin{abstract}

We study the fluctuations of the Gaussian model, with conservation of
the order parameter, evolving in contact with a thermal bath quenched from inverse temperature
$\beta _i$ to a final one $\beta _f$. At every time there exists a critical value $s_c(t)$ 
of the variance $s$ of the order parameter per degree of freedom
such that the fluctuations with $s>s_c(t)$ are characterized by
a macroscopic contribution of the zero wavevector mode, similarly to what occurs in an ordinary
condensation transition. We show that the probability of fluctuations with $s<\inf_t [s_c(t)]$, for which
condensation never occurs, rapidly converges towards a stationary behavior.  
By contrast, the process of populating the zero wavevector mode of the variance, which takes place 
for $s>\inf _t [s_c(t)]$, induces a slow non-equilibrium dynamics resembling that
of systems quenched across a phase transition.

\end{abstract}

\submitto{Journal of Statistical Mechanics: theory and experiment}

\maketitle

\section{Introduction} \label{intro}

The theory of large deviations deals with the probability of observing atypical and largely improbable 
events in statistical systems. Fundamental results in this branch of probability theory bear important 
consequences in several fields of science~\cite{Hinrichsen00,Langer92,Touchette2009}
and are successfully applied to various practical situations~\cite{Cramer38,Cramer44}. Under general conditions,
the probability $P(S)$ to observe a certain
value $S$ of a collective variable obeys a large deviation principle~\cite{Touchette2009}, i.e., $P(S)\sim e^{-VI(s)}$, where $V$
is a measure of the number of degrees of freedom contributing to $S$, assumed to be large, $s = S/V$ is the intensive variable associated with $S$,
and $I(s)$ the so-called {\it rate function} which is non-negative 
and it generically vanishes at the average and most probable value of $s$. 
The above holds in the large $V$ limit. 
For example, $S$ could be thought of as being an extensive 
macroscopic variable, such as the number of particles in an open system of large volume $V$ at a certain chemical potential and therefore $s$ is the particle density in that volume. 

If the conditions for the applicability of the central limit theorem are verified, small fluctuations of $S$ of order $\sqrt{V}$
are Gaussianly distributed around its average $\langle S \rangle = \langle s \rangle V$ and hence $I(s)$ is quadratic around the most probable outcome 
$\langle s\rangle$ of $s$. On the other hand, the large deviation principle describes the rare fluctuations of $S$ of order $V$ which are exponentially suppressed as $V$ increases and which can display a wealth of different and interesting behaviors. 
Notably, $P(S)$ can exhibit singular points~\cite{Corberi19,Baek_2015,Filiasi_2014,Harris_2009,Gradenigo_2013,Gambassi2012,2019arXiv190406259P,Goold2018,Touchette2007,Touchette_2009,Bouchet_2012,Harris_2005,Szavits2014,Chleboun2010,Janas2016,Sasorov_2017,Majumdar_2014} at which some derivatives are discontinuous. This fact is usually interpreted as a phase transition occurring at the level of fluctuating
configurations. Namely, if $s_c$ is one of these singular points, the configurations of the system corresponding
to $s<s_c$ or to $s>s_c$ are qualitatively different. This is exactly what occurs when an ordinary
phase transition is present in a statistical system. The difference is that in the latter case the typical and statistical properties of the system
change qualitatively when a control parameter (the role of which is played here by $s$) crosses a critical value 
(the analogous of $s_c$), whereas here there is no need to change any external parameter, because rare fluctuation spontaneously
occurring with $s<s_c$ or $s>s_c$ naturally correspond to radically different system properties.

In spite of the fact that large deviation theory has been widely used for studying the stationary properties of both equilibrium and non-equilibrium stochastic processes~\cite{Touchette2009}, the topic of the dynamics of large fluctuations is 
largely unexplored. The most general problem consists in understanding how an atypical state which realizes a rare fluctuation can be 
reached by the system starting from a certain, specified condition. A concrete example is that of two identical containers of total volume $V$ each containing 
a number $N$ of molecules of a gas, in the same thermodynamic conditions. If at some time $t=0$ they are connected
by a pipe which allows the exchange of particles between the two containers, the objective is to find the probability to observe
an improbable number $S\gg \langle S\rangle =N $, e.g., $S \simeq 1.5 N$, of particles in one of the two containers after an elapsed 
time $t$.     
 
In a previous paper~\cite{Corberi17}, this issue was addressed in a solvable model where 
$S=\sum _{k=1}^V s_k$ is the sum of a large number $V$ of independent and identically distributed 
variables $s_k$, which evolve in time according to a certain stochastic dynamics. Depending on the actual distribution of the $s_k$, the probability $P(S)$ 
can exhibit a singular point $S_c$. 
Starting from a typical state with $S=\langle S\rangle$, the probability $P(S,t)$ of finding any value $S$ was determined.
It was observed that the evolution of $P(S,t)$ is radically different if a critical point $S_c$ for the variable $S$ is
present or not. In its absence, $P(S,t)$ evolves quite smoothly and, in a relatively
short time, rare fluctuations with $S - \langle S\rangle \sim {\mathcal O}(V)$ are developed such that the probability
to observe them quickly attain its stationary value. If, instead, a critical point $S_c$ is present,
the evolution occurs as described above only on one side of the value $S_c$ (in that concrete example for $S<S_c$),
whereas on the other side, the evolution of $P(S,t)$ is slow and characterized by a never-ending
algebraic relaxation which strongly resembles the one observed in thermodynamic systems
brought across a phase transition~\cite{Bray94,CorCugYos11,Cor15,Corberi11}.
This fact reinforces the interpretation of a singular point in $P(S,t)$ as a sort of a phase transition.   

In this paper we study the dynamics of fluctuations in a prototypical model of statistical mechanics, i.e.,
the Gaussian model. 
In this system, 
the probability distribution of the variance $S$ 
of the order parameter displays a critical point $S_c$ both in and out of
equilibrium~\cite{Zannetti14,CORBERI2015,Corberi17,Zannetti14,Corberi19,Cagnetta17,Corberi_2015,Zannetti_2014,Corberi_2013,Corberi_2012}, 
where the model
experiences a condensation transition at the level of fluctuations, a phenomenon which has been termed
{\it condensation of fluctuations}. Accordingly, this is a natural candidate to study how the presence
of such a singularity affects the dynamical properties of large deviations, similarly to what was done
in Ref.~\cite{Corberi17}. 

We study here 
the evolution of $P(S,t)$ when the system is initially prepared in an equilibrium configuration at a certain temperature $\beta_i^{-1}$
and is subject at time $t=0$ to a quench, i.e., it is let subsequently to evolve with a dynamics corresponding to a different temperature $\beta_f^{-1}$. Differently from the case considered in Ref.~\cite{Corberi17}, with this
protocol, large deviations associated with condensed states of the system are present at any time. However, due to the abrupt change in the thermal conditions,
a non-condensed configuration associated with a certain value of $S$ can happen to cross $S_c$ during its non-equilibrium
evolution. Interpreting $S_c$ as a critical point, also such crossing represents the 
occurrence of a phase transition and therefore we expect it to result into a complex and slow kinetics,
as discussed above. By solving exactly the evolution equations of the model 
we show that this is actually what happens. Specifically, the evolution is trivial and quasi-adiabatical
for fluctuations  associated with a value $S$ which does not cross $S_c$ during the temporal evolution, while it is 
much richer and slow if it does. 

This paper is organized as follows: In Sec.~\ref{themodel} we introduce the Gaussian model and
its dynamics, considering in particular the quench protocol.
In Sec.~\ref{fluctuations} we determine the probability $P(S,t)$ and,
in Sec.~\ref{condensation} we discuss the condensation transition.
Section~\ref{dynamics} presents the main results concerning the evolution of $P(S,t)$, which is discussed in detail. Finally,
we draw our conclusions and highlight some additional questions and open points
in Sec.~\ref{conclusions}.

\section{The model} \label{themodel}

We consider the Gaussian model~\cite{Goldenfeld92,chaikin_lubensky_1995}, describing a scalar and real field
$\varphi (\vec x)$ (an order parameter in the language of phase transitions) the equilibrium properties of which are
governed by a Hamiltonian in $d$ dimensions
\be
   {\cal H}[\varphi]=\frac{1}{2}\int _V d\vec x \left [(\nabla \varphi )^2
     +r\varphi ^2(\vec x)\right ]=
   \sum _{\vec k}{\cal H}_{\vec k},
   \label{ham}
\ee
where $r\ge 0$ is the parameter which controls the extent $\xi = r^{-1/2}$ of the spatial correlations of the field in equilibrium, corresponding to criticality at $r=0$. On the r.h.s.,
\be
   {\cal H}_{\vec k}=
   \frac{1}{2V}\omega _k\varphi _{\vec k}\varphi_{-\vec k},
   \label{ham_k}
\ee
describes ${\cal H}$
in terms of the Fourier components
$\varphi _{\vec k}$ of the order parameter, 
where $\omega _k=k^2+r$, and $V$ is the volume occupied by the system.
Because of the fineteness of the volume the modes are quantized, therefore the sum on the r.h.s. of Eq.~(1), and
although the choice of boundary conditions is inconsequential in the present problem, we assume them to be periodic.
We further impose an ultraviolet cut-off $\Lambda$ accounting for a microscopic length scale, due for example to a lattice spacing, such that the allowed modes are all those with wavevectors of magnitude smaller than the cut-off.

Notice that reality of the order parameter field implies that
only one half of its Fourier components are independent, i.e., that $\varphi_{-\vec{k}} = \varphi_{\vec{k}}^*$.
We take this into account by letting $\vec k$ in $\sum_{\vec k}$ take values only on one half of the $\vec k$ space and multiplying by a factor 2. Accordingly, ${\mathcal H}_{\vec{k}}$ in Eq.~(2) is replaced by
\be
   {\cal H}_{\vec k}=
   \frac{1}{V}\chi _{k}\omega _k\varphi _{\vec k}\varphi_{-\vec k},
\ee
where we introduced the function \(\chi_{k}\), such that \(\chi_0=1/2\) and \(\chi_{k}=1\) otherwise,
which avoids counting twice the zero mode.

The dynamics of the model with local conservation of the order parameter is
given by the following overdamped Langevin evolution~\cite{Goldenfeld92,Hohenberg77}
\be
\frac {\partial \varphi(\vec x,t)}{\partial t}=
-\nabla ^2 \left [\nabla ^2 -r\right ]\varphi (\vec x,t)+\eta(\vec x,t),
\ee
where $\eta (\vec x,t)$ is assumed to be an uncorrelated Gaussian noise of thermal
origin, at temperature $\beta^{-1}$, with zero average and
\be
\langle \eta (\vec x,t)\eta(\vec x',t')\rangle=-2\beta^{-1}
\nabla^2\delta(\vec x -\vec x')\delta(t-t').
\ee
With this choice of dynamics, the stationary probability distribution function of the fluctuating field is generically an equilibrium one and is given by $P_{\rm eq}[\varphi] \propto {\rm e}^{- \beta {\mathcal H}[\varphi]}$. In Fourier space one has
\be
\frac {\partial \varphi _{\vec k}(t)}{\partial t}=
-\widetilde \omega _k\varphi _{\vec k}(t)+\eta_{\vec k}(t),
\label{evolutionk}
\ee
with $\widetilde \omega _k =k^2(k^2+r)$ and where the noise correlator is
\be
\langle \eta_{\vec{k}} (t)\eta_{\vec{k}'} (t')\rangle=\frac{V}{\chi_{k}}
\beta^{-1}k^2\delta_{\vec{k},-\vec{k}'}\delta(t-t').
\ee

In the following we will consider the dynamics induced by a sudden temperature
quench form an initial inverse temperature value $\beta_i =(k_B T_i)^{-1}$
($k_B$ being the Boltzmann constant), to a final one
$\beta _f>\beta _i$, operated at $t=0$.
We emphasize here that the choice of a final temperature larger than the initial one, i.e., \(\beta_f<\beta_i\),
leads to a different phenomenology compared to that discussed further below, which deserves a separate discussion beyond the scope of the present work.

The explicit solution of the evolution equation (\ref{evolutionk}) of each mode reads, for $t\ge 0$,
\be
\varphi_{\vec{k}}(t)=\varphi_{\vec{k}}(0)e^{-\tilde\omega_{k}t}+\int_0^tdt'\ e^{-\tilde\omega_k(t-t')}\zeta_{\vec{k}}(t').
\label{eq:solution_k-modes}
\ee
The correlation of the fields is therefore
\be
\langle \varphi_{\vec{k}}(t)\varphi_{-\vec{k}}(t) \rangle =\langle\varphi_{\vec{k}}(0)\varphi_{-\vec{k}}(0)\rangle_0 e^{-2\tilde\omega_k t}+\frac{\beta_f^{-1} V}
        {2\chi_{k}\omega_k}(1-e^{-2\tilde\omega_k t}),
\ee
where \(\langle...\rangle_0\) stands for the average over initial conditions. This implies that the instantaneous expectation value of the Hamiltonian is
\be
2\langle {\cal H}_{\vec k}\rangle =
\beta _k^{-1}(t)=\left (\beta _i^{-1}-\beta _f^{-1}\right )
e^{-2\widetilde \omega _k t}+\beta _f^{-1},
\label{defBeta}
\ee
where $\beta_k(t)$ has the heuristic meaning of a mode-dependent instantaneous non-equilibrium inverse temperature which interpolates between the initial $k$-independent value $\beta_k(0) = \beta_i$ and the final one $\beta_{k\neq 0}(t\to \infty) = \beta_f$,  determined by the equipartition theorem in equilibrium conditions.
During the non-equilibrium evolution, the equipartition theorem does not hold and, in fact, the expectation value $\langle {\mathcal H}_{\vec{k}}\rangle$ is not related to any temperature and it is mode-dependent.
Note that the effective
temperature $\beta^{-1}_k(t)$ is not necessarily a positive quantity and that
its value at $k=0$ is fixed by the initial condition due
to the conservation law of the order parameter (and therefore we will write 
$\beta _0$ instead of $\beta _0(t)$ in the following).

\section{Fluctuations of the variance} \label{fluctuations}

We will study the fluctuations of the order parameter variance
\be
   {\cal S}[\varphi]=\int _V d\vec x \,\varphi ^2(\vec x,t)=\frac{2}{V}
   \sum _{\vec k}\chi_{k}\varphi _{\vec k}(t)\varphi _{-\vec k}(t).
\label{defS}
\ee
The probability distribution of the value $S$ of this quantity
reads
\be
P(S,t)=\int _\Gamma D\varphi \, P([\varphi],t)\,\delta (S-{\cal S}[\varphi]),
\label{defPS}
\ee
where $\Gamma $ is the space of configurations of the field $\varphi$, $P([\varphi],t)$ is the probability
of one of such configurations at time $t$, and $\delta $ is the Dirac delta function.

In equilibrium conditions at inverse temperature $\beta $ one has
$P([\varphi],t)=P_{eq}([\varphi])=Z^{-1}e^{-\beta {\cal H}[\varphi]}$,
where $Z$ is the normalization constant.
It is easy to show that, 
considering equilibrium states at different
temperatures, one has (see \ref{app1})
\be
P_{eq}(S)=f\left ( \frac{S}{\langle S\rangle}\right ),
\label{scaleq}
\ee
where $\langle S \rangle =\int _0 ^\infty dS \, S\, P(S)=\beta ^{-1}\sum _{\vec k} \omega _k ^{-1}$, is the average value of $S$.
The scaling property~(\ref{scaleq}) means that the only effect
on $P_{eq}(S)$ of considering different temperatures
is to set a different scale $\langle S \rangle$ of $S$. Accordingly, by
measuring $S$ in units of $\langle S \rangle$ one recovers
the same universal behavior described by the function $f$ reported in Eq.~(\ref{scaleq}).

Because the problem is diagonalized in Fourier components, the phase-space
measure $P([\varphi],t)=\Pi _{\vec k} P_{\vec k}(\varphi_{\vec k},t)$ is factorized
at all times.
On the basis of the explicit solution for the field at a certain time given in Eq.~(\ref{eq:solution_k-modes}), it follows that the distribution of the single $\varphi_{\vec{k}}$ are Gaussian and therefore they are completely characterized by their (vanishing) average and variance, the latter being essentially encoded in ${\mathcal H}_{\vec{k}}$, the expectation value of which is reported in Eq.~(\ref{defBeta}).
Thus
\be
P_{\vec k}(\varphi_{\vec k},t)=
Z_{\vec k}^{-1}(t)e^{-\beta_k(t){\cal H}_{\vec k}(\varphi _{\vec k})},
\ee
where $Z^{-1}_{\vec k}(t)=\left [\frac
  {\chi_{k}\beta _k(t)\omega _k}{\pi V}\right ]^{\frac{1}{2}}$.

Expressing the $\delta$ function constraint in Eq.~(\ref{defPS}) via the
representation
$\delta (y)=\frac{1}{2\pi i}\int _{a-i\infty}^{a+i\infty} dz\, e^{-zy}$
one arrives at
\be
P(S,t)=\frac{1}{2\pi i}\int _{a-i\infty}^{a+i\infty} dz\,
e^{-V\left [zs+\lambda (z,t)\right ]},
\ee
where $s=S/V$ is the intensive variable associated with $S$, and
\be
\lambda(z,t)=
-\frac{1}{V}\ln \int D\varphi P([\varphi],t)e^{z{\cal S}[\varphi]}=-\frac{1}{V}\sum_{\vec{k}}\ln\frac{1}{\sqrt{1-\frac{2z}{\beta_k(t)\omega_k}}}
\label{lambda}
\ee
is the scaled cumulant generating function. 
In Eq. (15), $a$ is any real number such that $\lambda(z,t)$ is analytic for ${\rm Re\,}z > a$.
Using G\"{a}rtner-Ellis theorem \cite{Touchette2009}, for a large volume $V \to \infty$ one arrives at
the large deviation form
\be
P(S,t)\sim e^{-VI(s,t)},
\label{largeDevPr}
\ee
where the rate function $I(s,t)$ is given by
\be
I(s,t)=z^*(s,t)s+\lambda(z^*(s,t),t),
\label{RateFunct}
\ee
where $z^*(s,t)$ is determined by the extremization condition
\be
\left .\frac{\partial \lambda (z,t)}{\partial z} \right \vert _{z=z^*(s,t)}+s=0.
\label{extr}
\ee

\section{Condensation} \label{condensation}

In the large volume limit, if the sums over the wavevector $\vec{k}$ can be
transformed into an integral according to
$\frac{1}{V}\sum _{\vec k}\dots\to \int \frac{d\vec k}{(2\pi)^d}\dots$, where $d$ is the number of spatial dimensions,
the extremal condition~(\ref{extr}) reads
\be
s=\Omega _d\int _0 ^\Lambda \frac{dk}{(2\pi)^d}\,
\frac{k^{d-1}}{\beta_k(t)\omega _k-2z^*},
\label{autoc}
\ee
where $\Omega _d=2\pi^{d/2}/\Gamma(d/2)$ is the $d$-dimensional
solid angle, $\beta_k(t)$ is ginven in Eq.~(\ref{defBeta}), and $\Gamma (\dots)$ the Euler function.
This equation has to be solved in order to determine $z^* = z^*(s,t)$.
Since $s$ is positive by definition, $z$ must be smaller than $\beta_0(t)\omega_0/2$,
because, given Eq.~(\ref{defBeta}), $\beta_0$ is the smallest among the 
$\beta_k(t)$ upon varying $k$.
Notice that in the scenario with \(\beta_f<\beta_i\), \(\beta_0\) is no longer the smallest
and this is the main reason why in this case the resulting dynamics is markedly different, as pointed out before. 
The integral on the r.h.s. of Eq.~(\ref{autoc}) diverges
in the limit $z\to \beta_0\omega_0/2$ if $d\le 2$,
while it is finite
for $d>2$. In the latter case the solution of Eq.~(\ref{autoc}) 
exists only for values of $s$ smaller than $s_c(t)$ defined by the condition
\be
s_c(t)=\Omega _d\int _0 ^\Lambda \frac{dk}{(2\pi)^d}\,
\frac{k^{d-1}}{\beta_k(t)\omega _k-\beta_0\omega_0}.
\label{defSc}
\ee
For $s>s_c(t)$, the solution requires a careful mathematical
treatment~\cite{CORBERI2015}. Alternatively, the solution can
also be found within an approach motivated and inspired by
what is known for the Bose-Einsten condensation:
One singles out the mode $k=0$ from the momentum sum, transforming the rest into an integral as before, thus
arriving at
\be
s=\frac{1}{V}\,s_0(s,t)+
\Omega _d\int _0 ^\Lambda \frac{dk}{(2\pi)^d}\,
\frac{k^{d-1}}{\beta_k(t)\omega _k-2z^*},
\label{autoc2}
\ee
instead of Eq.~(\ref{autoc}), with
\be
s_0(s,t)=\frac{1}{\beta_0\omega_0-2z^*(s,t)}.
\label{defs0}
\ee
For $s< s_c(t)$,b one has
$z^*(s,t)\le \beta_0\omega_0/2 $ and hence the first term is negligible
for large $V$. For $s\ge s_c(t)$, instead, one has $z^*\equiv \beta_0\omega _0/2$ and,
for $s>s_c(t)$ this term becomes macroscopically large and takes the
value $s-s_c(t)$. As a consequence, the large deviation form~(\ref{largeDevPr})
holds with
\be
I(s,t)=\left \{ \begin{array}{ll}
z^*(s,t)s+\lambda(z^*(s,t),t) &  \mbox{for}\,\,\,\,\, s\le s_c(t) , \\
\beta_0\omega_0 (s-s_c)/2+I(s_c,t) &  \mbox{for}\,\,\,\,\, s>s_c(t) ,
\end{array}
\right .
\label{RateCase}
\ee
instead of Eq.~(\ref{extr}).
Because $I(s,t)$ is linear for $s \ge s_c(t)$ while it is not for $s \le s_c(t)$, the left and right derivatives with respect to $s$ at $s = s_c(t)$ differ at a certain order, lager than the first one \cite{CORBERI2015}.
Notice also that \(\lim_{s\to 0}I(s,t)=\infty\)\footnote{%
Indeed it can be easily checked from Eq.~(\ref{autoc}) that \(z^*(s,t)\to-\infty\) with \(z^*(s,t)s\to\) const.
and that \(\lim_{s\to0}\lambda(z^*(s,t),t)=\infty\), after Eq.~(\ref{lambda}).},
hence \(P(S=0,t)=0\), because \(S=0\) can be realized by the sole configuration \(\varphi\equiv0\). 

\section{Dynamics of fluctuations} \label{dynamics}

\begin{figure}[h]
\vspace{1.5cm}
  \centering
  \rotatebox{-90}{\resizebox{.7\textwidth}{!}{\includegraphics{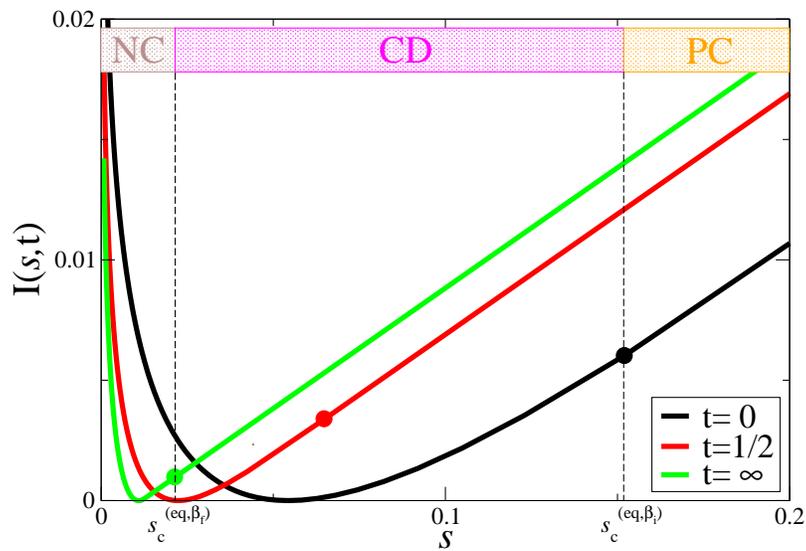}}}
  \caption{Rate function $I(s,t)$ as a function of $s$ for three different values of the time
   $t = 0$, $0.5$, and $\infty$ elapsed from the quench with $\beta _i=1/5$
    and $\beta _f=1$, in the case $d=3$ (the same qualitative features are observed for other values of $d>2$), with $r=1$, while the value of the ultraviolet cut-off $\Lambda$ is set to 1.
    The critical value $s_c(t)$ of the variable $s$ is marked by a thick dot. 
    The three regions NC, CD, and PC, discussed in the main text, are
    highlighted at the top of the figure.}
\label{fig1}
\end{figure}

In the following we will study the dynamics of the fluctuations
after the quench of the inverse temperature $\beta$ of the stochastic noise from $\beta _i$ to $\beta _f$.
The evolution of $I(s,t)$, in the sample case $d=3$, is shown in Fig.~\ref{fig1} for three different values of times,
i.e.,  $t=0$, corresponding to the initial state, $t=0.5$ and $t=\infty$, the latter corresponding to the eventual stationary state. 
According to the large deviation form~(\ref{largeDevPr}), the average
value $\langle s(t)\rangle$ corresponds to the minimum, which is also the zero, of $I(s,t)$
and its expression derives from Eq.~(\ref{autoc}) taking in account the fact that,
for the average, $z^*$ in~(\ref{RateFunct})vanish at all times
\begin{equation}
\langle s(t)\rangle=\frac{\Omega_d}{(2\pi)^d}\int_{0}^{\Lambda}dk\,\frac{k^{d-1}}{\beta_k(t)\omega_k}.
\label{AvgDef}
\end{equation}
Since the fluctuations of the order parameter are due to
thermal fluctuations, their typical value
$\langle s(t)\rangle$ moves from the
initial to the final equilibrium values
$\langle s\rangle^{(eq,\beta_i)}$, $\langle s\rangle^{(eq,\beta_f)}$,
obtained taking respectively $t=0$ and $t\rightarrow \infty$ in~(\ref{AvgDef}),
decreasing in time being $\beta_f>\beta_i$.
Using the model equation it can be shown (see \ref{app2})
that the evolution of the average variance for sufficiently long times is
\be
\langle s(t)\rangle=\langle s\rangle ^{(eq,\beta_f)}+A\ t^{-d/2},
\label{dynAveS}
\ee
with
\begin{equation}
A=\frac{\Omega_d(\beta_i^{-1}-\beta_f^{-1})\Gamma(d/2)}{r(2\pi)^d(2r)^{d/2+1}},
\end{equation}
where \(\Gamma\) is the Gamma function.
Because of the Gaussian nature of the problem the critical
point $s_c(t)$, above which condensation occurs,
must also decrease proportionally to what
$\langle s \rangle$ does. Solving the model equations (see \ref{app3})
one finds that during the non-equilibrium evolution $s_c(t)$ decreases
monotonically and, at long times, one has
\be
s_c(t)=s _c^{(eq,\beta_f)}+a\ t^{-d/2},
\label{dynSc}
\ee
with 
\begin{equation}
a=\frac{\Omega_d\beta_{f}\Gamma(d/2)\zeta(d/2)}{(2\pi)^d(2r)^{d/2+1}\beta_i(\beta_f-\beta_i)},
\end{equation}
where $\zeta$ is the Riemann zeta function.

During the process, the slope $\beta _0\omega _0/2$ of the linear branch of $I(s,t)$
corresponding to condensation (see Eq.~(\ref{RateCase})) is fixed
because, as already observed, $\beta _0$ is time-independent. This means that,
in the condensed region for $s>s_c(\infty)$, the rate function
$I(s,\infty)$ at $t=\infty$ cannot be superimposed on the initial one $I(s,0)$ using
the equilibrium relation~(\ref{scaleq}). The reason of this apparent incongruence is that the two equilibrium 
states are different not only because of $\beta $, as implicit in Eq.~(\ref{scaleq}), but also
because of the reduction of the set of possible final states that can be reached, starting from an assigned inital one,
by the conserved dynamics.

Given this phenomenology, it is clear that the evolution of $I(s,t)$
displays different features depending on whether condensation occurs or not.
Indeed we argue below that the dynamical process accompanying condensation,
namely the building up of a macroscopic $s_0(t)$ out of a microscopic
initial value $s_0(t=0)$, is much slower and collective
than the easier rearrangement of fluctuations occurring at values of $s$ for which this does not occur.
On the basis of these considerations we can divide
the range of values of $s$ into three different within which fluctuations have markedly
different character, as also indicated in Fig.~\ref{fig1}.

\subsection{Non-condensed (NC) region}

This region corresponds to $s<s_c^{(eq,\beta_f)}= s_c(t=\infty)$ and is characterized by the fact that condensation never occurs during the dynamics and all the fluctuating modes $s_k$ contribute to the final value $s = \sum_{\vec{k}} s_k$ of the variance with ``microscopic'' contributions of order $1/V$. Accordingly, during the dynamics,
one simply observes the redistribution of their contributions 
in order for the fluctuations to pass smoothly from the initial to the
final equilibrium behaviors.
Give that such a redistribution involves only modes which provide microscopic contributions -- contrary to what happens when condensation occurs -- we expect the dynamics within this NC region to be fast. 

\begin{figure}[h]
\vspace{1.5cm}
  \centering
\rotatebox{-90}{\resizebox{.7\textwidth}{!}{\includegraphics{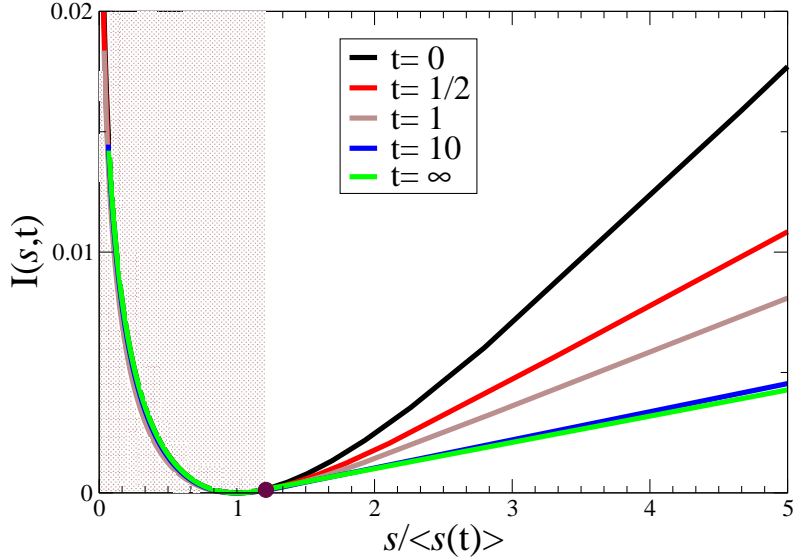}}}
  \caption{Rate function $I(s,t)$ as a function of the rescaled variable $s/\langle s(t)\rangle$
    for various fixed values of the time $t$ after  a quench from $\beta _i=1/5$
    to $\beta _f=1$, with $r=1$, $\Lambda=1$, in the case $d=3$.
    The critical value $s_c^{(eq,\beta _f)}/\langle s\rangle ^{(eq,\beta_f)}$ is marked by a
    thick dot and the NC region is highlighted by a brown background.}
\label{fig2}
\end{figure}

We rationalise this hypoteses as follows: in a system at equilibrium,
the scaling in Eq.~(\ref{scaleq}) holds true. Clearly, the same does not
hold {\it a priori} out of equilibrium and, indeed, there is no way to
show it as one does in the case of equilibrium discussed in \ref{app1}.
However, if the process of rearrangement 
occurs {\it quasi adiabatically}, we would expect the only effect
of the quench on $I(s,t)$ to be the shift of $\langle s(t)\rangle $,
according to Eq.~(\ref{dynAveS}), without affecting the form of $f(y)$ reported in
Eq.~(\ref{scaleq}). In this case, plotting $I(s,t)$ for a fixed time $t$ as a function of 
$s/\langle s(t)\rangle $, one should observe superposition
of the curves at different times on the mastercurve $f(y)$, formally corresponding to the case $t=\infty$.
This scenario is tested in Fig.~\ref{fig2}, where one clearly sees that in
the NC region (namely to the left of the thick dot in the figure) 
curves corresponding to different times superimpose almost perfectly at all times, implying an adiabatic evolution.

Clearly, the scaling encoded in Eq.~(\ref{scaleq}) and observed in the NC region
is not expected to be exact, as in equilibrium, but it anyhow turns out to be an excellent
approximation. In particular, Eq.~(\ref{scaleq}) does not hold out of equilibrium because
now in Eq.~(\ref{scalneq}) there is an explicit time dependence
in $\langle \psi _{\vec k}(t)\psi _{-\vec k}(t)\rangle$, where $\psi_{\vec{k}}=\langle s(t)\rangle^{-\frac{1}{2}}\,\varphi _{\vec k}$ is the rescaled field (see \ref{app1}). The observed approximate scaling behavior
might be possibly due to the fact that the domain of integration in Eq.~(\ref{scalneq}), given by the part of $\Gamma $ where
the argument of the $\delta$-function vanishes,
for $s<s_c$ constrains the integration variables
$\psi _{\vec k}$ in regions much smaller than their variances
$\omega _k^{-1}\langle \psi _{\vec k}(t)\psi _{\vec k}(t)\rangle $, thereby
making the time-dependence induced by the dynamics largely irrelevant.
Clearly this is not possible in the presence of condensation since
the variance of the mode with $k=0$ grows macroscopic. 

\subsection{Condensation-developing (CD) region}

\begin{figure}[h]
\vspace{1.5cm}
  \centering
\rotatebox{-90}{\resizebox{.7\textwidth}{!}{\includegraphics{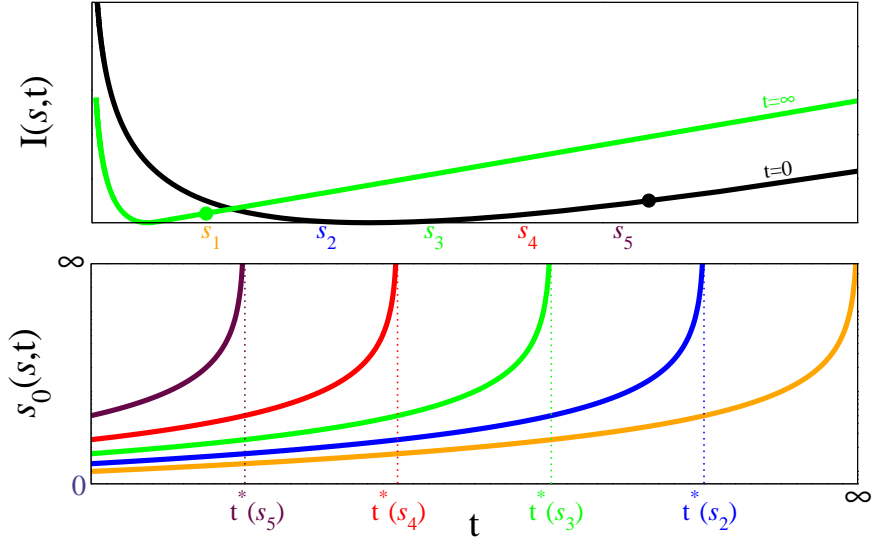}}}
  \caption{Upper panel: Rate function $I(s,t)$ as a function of $s$ 
   the initial equilibrium state , i.e., immediately before the quench (black line, $t=0$) and in the final one ($t=\infty$).
   In both cases, the corresponding critical value of $s$ are indicated by dots.
   Lower panel: time dependence of $s_0(s,t)$, for various values ($s_1,s_2,s_3,s_4,s_5$) of $s$
   which, for comparison, are located in the upper panel with respect to the rate functions at $t=0$ and $t=\infty$.
   Both panels refer to the case $d=3$, $r=1$ and $\Lambda = 1$ for a quench from $\beta _i=1/5$
   to $\beta _f=1$.}
\label{fig3}
\end{figure}

Any fixed value of $s$ within the CD region $s_c^{(eq,\beta_f)}< s< s_c^{(eq,\beta_i)}$
is crossed by $s_c(t)$ at a certain time $t^*(s)$. Note that $s_c^{(eq,\beta_f)}< s_c^{(eq,\beta_i)}$
holds due to Eq.~(\ref{defSc}).
This implies that for $t<t^*(s)$, the contribution to the average variance 
of the zero mode $s_0(s,t)$ is a finite quantity in the thermodynamic limit, with $s_0 \sim {\mathcal O}(V^0)$.
Instead, for $t>t^*(s)$, $s_0(s,t)$ diverges in the same limit, with $s_0 \sim {\mathcal O}(V)$.
This is pictorially sketched in Fig.~\ref{fig3}, 
where in the lower panel the time behavior of $s_0(s,t)$ for various values 
($s_1,s_2,s_3,s_4,s_5$) of $s$ within the CD region is shown; 
in the upper panel of the same figure the position of these values is shown in relationship with the rate functions at $t=0$ and at $t=\infty$, the critical values of which (indicated by the dots) define the boundaries of the CD region.
For times $t\lesssim t^*(s)$ the divergence of $s_0(s,t)$ occurs as (see \ref{app4})
\be
\lim _{V\to \infty}s_0(s,t) \simeq \left \{ \begin{array}{ll}
[t^*(s)-t]^{-1} & \mbox{for} \,\,\,\,\, s>s_c^{(eq,\beta _f)}, \\
t^{d/2} & \mbox{for} \,\,\,\,\, s=s_c^{(eq,\beta _f)}, 
\end{array}
\right .
\label{divS0}
\ee
i.e., $s_0(s,t)$ with $s>s_c^{(eq,\beta _f)}$ diverges linearly while $s_0(s_c^{(eq,\beta _f)},t)$ algebraically.

Figure~\ref{fig2} shows that the relaxation of the
rate function in the CD region is much slower than that in the NC region. Indeed, while for the times
reported in the figure fluctuations are adiabatically in equilibrium in the NC region 
(corresponding to the values of $s$ on the left of the thick dot), in the CD and in the further region denoted as PC
(discussed below) where condensation is present from the beginning (on the right of 
the dot) a significant change is observed and convergence occurs only at much longer
times ($t\gtrsim 10$ on the scale of the present figure). 
 
\subsection{Permanent-condensation (PC) region}

For $s > s_c^{(eq,\beta_i)}$, $s_0(s,t)/V$ increases monotonically in time from 
$s-s_c^{(eq,\beta _i)}$ to $s-s_c^{(eq,\beta _f)}$. Also in this case
$s_0(s,t)$ changes by an infinite amount in the thermodynamic limit $V\to\infty$.
This is similar to what happens when the value of $s$ is within the CD region described above,
apart from the fact that in the latter case $s_0(s,0)$ is finite. 
Accordingly, we observe also in this case that fluctuations
do not relax adiabatically. Notice also that, no matter how large $t$ is, for sufficiently large values
of $s$ the rate function $I(s,t)$ differs significantly from its asymptotic form. A similar behavior was observed
in Ref.~\cite{Corberi17}.

\section{Conclusions} \label{conclusions}

In this paper we have analysed some aspects of the dynamics of fluctuations of
the variance $s$ per degree of freedom, of the order parameter in the Gaussian model with a conserved stochastic dynamics,
in which large deviations
may display the phenomenon of condensation. After a quench of the temperature
of the thermal bath the model is in contact with, 
we have shown that the non-equilibrium behavior of fluctuations is radically different 
depending on whether the selected value of $s$ is affected or not by the condensation as time goes by.
In particular, fluctations
which do not condense converge almost adiabatically to a stationary, equilibrium-like form.
Those affected by the condensation, instead, display a slow and complex evolution determined
by the slow contribution $s_0(s,t)$ of the $k=0$ wavevector. 

The emergence of these two qualitatively different behaviors, which was already
observed in another solvable model~\cite{Corberi17} of statistical mechanics, has a nice 
interpretation in the framework of what is known for ordinary phase transitions. It must be
recalled, in fact, that the expression~(\ref{defS}) of the probability we consider is formally
equivalent~\cite{Zannetti14,Touchette2009} to the partition function of a Gaussian model on a 
reduced phase space where the order-parameter variance is fixed to take the value $S$. 
This correspondence is usually referred to as \textit{duality}. A well-known model with such a constraint is the
spherical model of Berlin and Kac~\cite{BerlinKac52}. This model has a ferromagnetic to paramagnetic 
phase transition located at $s_c(\beta )$. Crossing a critical point in magnetic models
induces a slow, never-ending (in the thermodynamic limit) coarsening phenomenon 
characterized by an algebraic growth of a quantity that sets the scale of spatial fluctuations.
Indeed, the zero wavevector mode of the structure factor diverges because of the formation of the 
Bragg peak, 
$\lim _{t\to \infty}\langle \varphi (\vec k,t)\varphi(-\vec k,t)\rangle \sim \delta (\vec k)$.
In the problem considered in this work,
the values of $s$ within the CD region are crossed, at a certain time, by $s_c(t)$ and therefore
they are expected to share some of the properties of the slow kinetics observed in quenched 
ferromagnets. In fact, we have shown that this is actually the case, and the quantity $s_0(s_c^{(eq,\beta_f)},t) $ diverges algebraically.
Clearly, relaxation in the NC region is much faster, corresponding --- according to the analogy drawn above --- 
to quenching a ferromagnetic system without crossing the critical point.

In the present work we focussed on a particular kind of quench, 
in which the temperature $\beta ^{-1}$ of the thermal bath responsible for the stochastic noise is changed abruptly.
It must be noticed that letting the initial value $\beta_i\to 0$ implies that $s_c^{(eq,\beta_i)}$ grows to infinity
and, accordingly, no condensation occurs in the initial state. This 
case, which is recovered as a special limit of the solution presented in this work, is more
closely related to what was done in Ref.~\cite{Corberi17} where condensation is initially absent as well. 

While we studied here the case of a quench of the temperature of the thermal bath, 
one might consider different kind of quenches, e.g., those in which other parameters are varied,
such as $r$ or, equivalently, the coefficient of the square gradient term in Eq.~(\ref{ham}) (which we fixed here to be one
for simplicity). Similarly, other observables beyond the order-parameter variance could be considered.
Apart from quantitative specific differences, we expect to observe in all these cases phenomena similar to those described here,
with markedly different behavior of fluctuations depending on whether they cross or not a critical point.
Analogously, they are expected in the Gaussian model with purely relaxational dynamics, i.e., without conservation of
the order parameter, with the notable difference that, in this case, 
the relaxation occurs exponentially fast in time, in contrast to the algebraic one observed in the present model (see for instance
Eqs.~(\ref{divS0}), (\ref{dynAveS}) and (\ref{dynSc})).

The model considered here, and the related cases discussed above, as well as the model 
considered in Ref.~\cite{Corberi17} are characterized by independently fluctuating modes. However, there are 
examples of probability distributions which display a behaviour similiar to the one discussed in this work
also in more complex systems in which these modes interact,
for instance in intrinsically non-equilibrium states of models of active matter~\cite{Cagnetta17,Nemoto19}. The dynamics of 
fluctuations in these cases is largely unexplored and represents an interesting topic for
further investigations.

\section*{Acknowledgments}

F.C. acknowledges funding from PRIN 2015K7KK8L.

\vspace{4cm}

\bibliographystyle{iopart-num}

%\nocite{*}
\bibliography{fluctuations}

\providecommand{\newblock}{}
\begin{thebibliography}{10}
\expandafter\ifx\csname url\endcsname\relax
  \def\url#1{{\tt #1}}\fi
\expandafter\ifx\csname urlprefix\endcsname\relax\def\urlprefix{URL }\fi
\providecommand{\eprint}[2][]{\url{#2}}

\bibitem{Hinrichsen00}
Hinrichsen H 2000 {\em Adv. Phys.\/} {\bf 49} 815

\bibitem{Langer92}
Langer J 1992 {\em Solids far from Equilibrium\/} (Cambridge: Cambridge
  University Press) pp 297--363

\bibitem{Touchette2009}
Touchette H 2009 {\em Phys. Rep.\/} {\bf 478} 1

\bibitem{Cramer38}
Cram\'er H 1938 {\em Colloque Consacr\'e \`a la Th\'eorie Des Probabilit\'es.
  Vol.3.\/} (Paris: Hermann)

\bibitem{Cramer44}
Cram\'er H 1944 {\em Usp. Mat. Nauk\/} {\bf 10} 166

\bibitem{Corberi19}
Corberi F and Sarracino A 2019 {\em Entropy\/} {\bf 21} 312

\bibitem{Baek_2015}
Baek Y and Kafri Y 2015 {\em J. Stat. Mech.\/} {\bf 2015} P08026

\bibitem{Filiasi_2014}
Filiasi M, Livan G, Marsili M, Peressi M, Vesselli E and Zarinelli E 2014 {\em
  J. Stat. Mech.\/} {\bf 2014} P09030

\bibitem{Harris_2009}
Harris R~J and Touchette H 2009 {\em J. Phys. A: Math. Theor.\/} {\bf 42}
  342001

\bibitem{Gradenigo_2013}
Gradenigo G, Sarracino A, Puglisi A and Touchette H 2013 {\em J. Phys. A: Math.
  Theor.\/} {\bf 46} 335002

\bibitem{Gambassi2012}
Gambassi A and Silva A 2012 {\em Phys. Rev. Lett.\/} {\bf 109} 250602

\bibitem{2019arXiv190406259P}
{Perfetto} G, {Piroli} L and {Gambassi} A 2019  arXiv:1904.06259

\bibitem{Goold2018}
Goold J, Plastina F, Gambassi A and Silva A 2018 {\em The Role of Quantum Work
  Statistics in Many-Body Physics\/} (Cham: Springer International Publishing)
  pp 317--336

\bibitem{Touchette2007}
Touchette H and Cohen E~G~D 2007 {\em Phys. Rev. E\/} {\bf 76} 020101

\bibitem{Touchette_2009}
Touchette H and Cohen E~G~D 2009 {\em Phys. Rev. E\/} {\bf 80} 011114

\bibitem{Bouchet_2012}
Bouchet F and Touchette H 2012 {\em J. Stat. Mech.\/} {\bf 2012} P05028

\bibitem{Harris_2005}
Harris R~J, R{\'{a}}kos A and Schütz G~M 2005 {\em J. Stat. Mech.\/} {\bf
  2005} P08003

\bibitem{Szavits2014}
Szavits-Nossan J, Evans M~R and Majumdar S~N 2014 {\em Phys. Rev. Lett.\/} {\bf
  112} 020602

\bibitem{Chleboun2010}
Chleboun P and Grosskinsky S 2010 {\em J. Stat. Phys.\/} {\bf 140} 846

\bibitem{Janas2016}
Janas M, Kamenev A and Meerson B 2016 {\em Phys. Rev. E\/} {\bf 94} 032133

\bibitem{Sasorov_2017}
Sasorov P, Meerson B and Prolhac S 2017 {\em J. Stat. Mech.\/} {\bf 2017}
  063203

\bibitem{Majumdar_2014}
Majumdar S~N and Schehr G 2014 {\em J. Stat. Mech.\/} {\bf 2014} P01012

\bibitem{Corberi17}
Corberi F 2017 {\em Phys. Rev. E\/} {\bf 95} 032136

\bibitem{Bray94}
Bray A 1994 {\em Adv. Phys.\/} {\bf 43} 357

\bibitem{CorCugYos11}
F~Corberi LF~Cugliandolo H~Y 2011 {\em Dynamical heterogeneities in glasses,
  colloids, and granular media\/} ed Berthier L, Biroli G, Bouchaud J~P,
  Cipelletti L and van Saarloos W (Oxford: Oxford University Press)

\bibitem{Cor15}
Corberi F 2015 {\em C. R. Phys.\/} {\bf 16} 332

\bibitem{Corberi11}
Corberi F, Lippiello E and Zannetti M 2002 {\em Phys. Rev. E\/} {\bf 65} 046136

\bibitem{Zannetti14}
Zannetti M, Corberi F and Gonnella G 2014 {\em Phys. Rev. E\/} {\bf 90} 012143

\bibitem{CORBERI2015}
Corberi F, Gonnella G and Piscitelli A 2015 {\em Journal of Non-Crystalline
  Solids\/} {\bf 407} 51

\bibitem{Cagnetta17}
Cagnetta F, Corberi F, Gonnella G and Suma A 2017 {\em Phys. Rev. Lett.\/} {\bf
  119} 158002

\bibitem{Corberi_2015}
Corberi F 2015 {\em J. Phys. A: Math. Theor.\/} {\bf 48} 465003

\bibitem{Zannetti_2014}
Zannetti M, Corberi F, Gonnella G and Piscitelli A 2014 {\em Commun. Theor.
  Phys.\/} {\bf 62} 555

\bibitem{Corberi_2013}
Corberi F, Gonnella G, Piscitelli A and Zannetti M 2013 {\em J. Phys. A: Math.
  Theor.\/} {\bf 46} 042001

\bibitem{Corberi_2012}
Corberi F and Cugliandolo L~F 2012 {\em J. Stat. Mech.\/} {\bf 2012} P11019

\bibitem{Goldenfeld92}
Goldenfeld N 1992 {\em Lectures on Phase Transitions and the Renormalization
  Group\/} (Reading: Addison-Wesley)

\bibitem{chaikin_lubensky_1995}
Chaikin P~M and Lubensky T~C 1995 {\em Principles of Condensed Matter
  Physics\/} (Cambridge: Cambridge University Press)

\bibitem{Hohenberg77}
Hohenberg P~C and Halperin B~I 1977 {\em Rev. Mod. Phys.\/} {\bf 49} 435

\bibitem{BerlinKac52}
Berlin T~H and Kac M 1952 {\em Phys. Rev.\/} {\bf 86} 821

\bibitem{Nemoto19}
Nemoto T, Fodor E, Cates M~E, Jack R~L and Tailleur J 2019 {\em Phys. Rev. E\/}
  {\bf 99} 022605

\end{thebibliography}

\vspace{1cm}

\appendix
\section{} \label{app1}

In this Appendix we prove the scaling property in Eq.~(\ref{scaleq}) for the equilibrium distribution function
$P_{eq}(S)$ of the variable $S$ in Eq.~(\ref{defS}).
Starting from Eq.~(\ref{defPS}) we change variable as 
$(\langle S(t)\rangle/V)^{1/2}$
one has
\be
\begin{array}{ll}
P(S,t)=&Z^{-1}(t)\int _\Gamma D\psi \, \exp \left \{-\frac{1}{2V}
  \sum _{\vec k} \omega _k\frac{\psi_{\vec k}\psi_{-\vec k}}
       {\langle \psi_{\vec k}(t)\psi_{-\vec k}(t)\rangle}\right \}\times \\
       &\delta \left (\frac {1}{V}\sum _{\vec k}\psi_{\vec k}\psi_{-\vec k}
       -\frac{S}{\langle S(t)\rangle}\right ).
\end{array}
\label{scalneq}
\ee       
In equilibrium all the time dependences drop out, and
$\langle \psi_{\vec k}\psi_{-\vec k}\rangle $
is independent of the temperature (and of $\vec k$),
due to the equipartition theorem.
Hence one has Eq.~(\ref{scaleq}).

\section{} \label{app2}

In this Appendix we derive the expression for the evolution at long times of the average of $s$. 
In the large-volume limit we have, from Eq.~(\ref{AvgDef}) 
\begin{equation}
\label{eq:average_COP}
\begin{array}{ll}
\langle s(t)\rangle&=\frac{\Omega_d}{(2\pi)^d}\int_{0}^{\Lambda}dk\,\frac{k^{d-1}}{\beta_k(t)\omega_k} \\
&=\frac{\Omega_d}{(2\pi)^d}\int_{0}^{\Lambda}dk\frac{k^{d-1}((\beta_i^{-1}-\beta_f^{-1})e^{-2k^2(k^2+r)t}+\beta_f^{-1})}{k^2+r}.
\end{array}
\end{equation}
The final equilibrium value, obtained for \(t\rightarrow\infty\) in the previous expression, reads
\begin{equation}
\label{eq:average_COP_eq}
\langle s\rangle^{(eq,\beta_f)}=\frac{\Omega_d}{(2\pi)^d}\int_{0}^{\Lambda}dk\,\frac{k^{d-1}}{\beta_f(k^2+r)}.
\end{equation}
The difference \(\langle s(t)\rangle-\langle s\rangle^{(eq,\beta_f)}\) is therefore given by
\begin{equation}
\langle s(t)\rangle-\langle s\rangle^{(eq,\beta_f)}=\frac{\Omega_d(\beta_i^{-1}-\beta_f^{-1})}{(2\pi)^d}\int_{0}^{\Lambda}dk\,\frac{k^{d-1}e^{-2k^2(k^2+r)t}}
{k^2+r}.
\end{equation}
Changing variable $x=t^{\frac{1}{2}} k$ leads to
\begin{equation}
\langle s(t)\rangle-\langle s\rangle^{(eq,\beta_f)}=\frac{\Omega_d(\beta_i^{-1}-\beta_f^{-1})t^{-d/2}}{(2\pi)^d}\int_{0}^{\Lambda\sqrt{t}}dx\,\frac{x^{-d/2}e^{-2x^2(x^2/t+r)}}
{x^2/t+r}.
\end{equation}
For large $t$, due to the fact that only small \(x\) contribute, the integral can be 
written as
\begin{equation}
\langle s(t)\rangle-\langle s\rangle^{(eq,\beta_f)}\simeq\frac{\Omega_d(\beta_i^{-1}-\beta_f^{-1})t^{-d/2}}{r(2\pi)^d}\int_{0}^{\infty}dx\,x^{-d/2}e^{-2x^2r}.
\end{equation}
Accordingly, one recovers Eq.~(\ref{dynAveS}), with
\begin{equation}
\begin{array}{ll}
A&=\frac{\Omega_d(\beta_i^{-1}-\beta_f^{-1})}{r(2\pi)^d}\int_{0}^{\infty}dx\,x^{-d/2}e^{-2x^2r}\\
&=\frac{\Omega_d(\beta_i^{-1}-\beta_f^{-1})\Gamma(d/2)}{r(2\pi)^d(2r)^{d/2+1}},
\end{array}
\end{equation}
where \(\Gamma\) is the Gamma function.
In order to assess the accuracy of this approximation for large $t$ we evaluated numerically $\langle s(t)\rangle$ in the case $d=3$, finding almost perfect correspondence. 

\section{} \label{app3}

In this Appendix we determine the evolution at large times of the critical point $s_c$ at long times, proceeding as in \ref{app2}. Here we have, again in the large-volume limit, starting from Eq.~(\ref{defSc})
\begin{equation}
\label{eq:criticalpoint_COP}
s_c(t)-s_c^{(eq,\beta_f)}=\frac{\Omega_d}{(2\pi)^d}\int_{0}^{\Lambda}dk\,k^{d-1}\Bigl[\frac{1}{(k^2+r)\beta_k(t)-r\beta_i}-\frac{1}{(k^2+r)\beta_f-r\beta_i}\Bigl],
\end{equation}
where \(\beta_k(t)\) is defined in~(\ref{defBeta}).
Changing variables $x=t^{\frac{1}{2}} k$ one has
\begin{equation}
s_c(t)-s_c^{(eq,\beta_f)}=\frac{\Omega_dt^{-d/2}}{(2\pi)^d}\int_{0}^{\Lambda\sqrt{t}}dx\,x^{d-1}\Bigl[\frac{1}{\frac{x^2/t+r}{(\beta_i^{-1}-\beta_f^{-1})e^{-2x^2(x^2/t+r)}+\beta_f^{-1}}-r\beta_i}-\frac{1}{(x^2/t+r)\beta_f-r\beta_i}\Bigl].
\end{equation}
For large $t$ we end up with
\begin{equation}
\label{eq:sc_appendix}
s_c(t)-s_c^{(eq,\beta_f)}\simeq\frac{\Omega_dt^{-d/2}}{(2\pi)^d}\int_{0}^{\infty}dx\,x^{d-1}\Bigl[\frac{1}{\frac{r}{(\beta_i^{-1}-\beta_f^{-1})e^{-2x^2r}+\beta_f^{-1}}-r\beta_i}-\frac{1}{r(\beta_f-\beta_i)}\Bigl],
\end{equation}
namely Eq.~(\ref{dynSc}).
The value of the coefficient \(a\) introduced therein is
\begin{equation}
\begin{array}{ll}
a&=\frac{\Omega_d}{(2\pi)^d}\int_{0}^{\infty}dx\,x^{d-1}\Bigl[\frac{1}{\frac{r}{(\beta_i^{-1}-\beta_f^{-1})e^{-2x^2r}+\beta_f^{-1}}-r\beta_i}-\frac{1}{r(\beta_f-\beta_i)}\Bigl] \\
&=\frac{\Omega_d\beta_{f}\Gamma(d/2)\zeta(d/2)}{(2\pi)^d(2r)^{d/2+1}\beta_i(\beta_f-\beta_i)},
\end{array}
\end{equation}
where \(\zeta\) is the Riemann zeta function.
Numerical calculations, for the test case $d=3$, confirm Eq.~(\ref{eq:sc_appendix}) with excellent accuracy.

\section{} \label{app4}

In this Appendix we determine the dynamics of the condensing mode $s_0(s,t)$
in the condensation-developing region.
We recall that $s_0(s,t)$ is defined in Eq.~(\ref{defs0}),
where \(z^{*}(s,t)\) is defined via Eq.~(\ref{extr})
\begin{equation}
s=\Omega_d\int_0^{\Lambda}\frac{dk}{(2\pi)^{d}}\,\frac{k^{d-1}}{\beta_k(t)\omega_k-2z^*},
\label{eq:s_appendix}
\end{equation}
with \(\omega_k\) defined after Eq.~(\ref{ham_k}).
In order to determine the behaviour of \(s_0(s,t)\) as time goes by, we must first determine that of \(z^*(s,t)\).
Taking the time derivative of Eq.~(\ref{eq:s_appendix}) one has
\begin{equation}
0=\int_0^{\Lambda}dkk^{d-1}\frac{[\tilde\omega_k\omega_k\beta_k^2(t)(\beta_i^{-1}-\beta_f^{-1})e^{-2\tilde\omega_kt}-\dot{z}^*(s,t)]}{[\beta_k(t)\omega_k-2z^*(s,t)]^2},
\end{equation}
where \(\dot z^*\) stands for the time derivative of \(z^*\), while \(\beta_k(t)\) is given in Eq.~(\ref{defBeta})
and \(\tilde\omega_k\) is defined after Eq.~(\ref{evolutionk}).
Accordingly,
\begin{equation}
\label{eq:startingpoint}
\dot z^*=\frac{\int_0^{\Lambda}dk\frac{k^{d-1}\tilde\omega_k\omega_k\beta_k^2(t)(\beta_i^{-1}-\beta_f^{-1})e^{-2\tilde\omega_kt}}{[\beta_k(t)\omega_k-2z^*(s,t)]^2}}
{\int_0^{\Lambda}dk\frac{k^{d-1}}{[\beta_k(t)\omega_k-2z^*(s,t)]^2}}.
\end{equation}
In order to proceed, we distinguish between values of $s$ inside the CD region, i.e.,
$s_c^{(eq,\beta_f)}<s<s_c^{(eq,\beta_i)}$ and the limiting value $s=s_c^{(eq,\beta_f)}$.
In the former case, \(s_0\) will diverge in a finite time so in the integrals in Eq.~(\ref{eq:startingpoint}) we can consider the limit of small \(k\), which also correspond to the portion of the domain where the variation in time is more important.
Accordingly, from Eq.~(\ref{eq:startingpoint}) one finds
\begin{equation}
\dot z^*=\frac{\int_0^{\Lambda}dk\,\frac{\frac{k^{d+1} r^2 (\beta_i^{-1}-\beta_f^{-1})}{(\beta_i^{-1})^2}}{\left(\frac{r}{\beta_i^{-1}}-2 z^*\right)^2}}
{\int_0^{\Lambda}dk\,\frac{k^{d-1}}{\left(\frac{r}{\beta_i^{-1}}-2 z^*\right)^2}}=\frac{\int_0^{\Lambda}dk\,\frac{k^{d+1} r^2 (\beta_i^{-1}-\beta_f^{-1})}{(\beta_i^{-1})^2}}
{\int_0^{\Lambda}dk\,k^{d-1}}=\mbox{const.}
\end{equation}
Thus \(z^*\) is linear in \(t\), implying
\begin{equation}
s_0(s,t)\simeq(t^*(s)-t)^{-1}, \qquad \forall s\in (s_c^{(eq,\beta_f)},s_c^{(eq,\beta_i)}).
\end{equation}
In the other case, \(s= s_c^{(eq,\beta_f)}\) is at the border of the CD region and
we can consider the limit of long times in Eq.~(\ref{eq:startingpoint}).
The integrand in the denominator, then, can be approximated by its leading behaviour
for small \(k\ll r^{1/2}\), i.e., as a constant 
\begin{equation}
\beta_k(t)\omega_k-2z^*(s_c^{(eq,\beta_f)},t)\simeq\beta_fr-2z^*(s_c^{(eq,\beta_f)},\infty).
\end{equation}
Accordingly, Eq.~(\ref{eq:startingpoint}) in the same limit renders
\begin{equation}
\dot z^*\simeq \frac{d}{\Lambda^d}\int_0^{\Lambda}dk\frac{k^{d-1}k^2r^2(\beta_i^{-1}-\beta_f^{-1})e^{-2k^2rt}}{[(\beta_i^{-1}-\beta_f^{-1})e^{-2k^2rt}+\beta_f^{-1}]^2},
\end{equation}
where we used that \(\tilde\omega_k\simeq k^2r\) and \(\omega_k\simeq r\).
The change of variables \(x=t^{\frac{1}{2}}k\) gives
\begin{equation}
\dot z^*\simeq \frac{t^{\frac{d+2}{2}}d}{\Lambda^d}\int_0^{\Lambda\sqrt{t}}dk\frac{x^{d+1}r^2(\beta_i^{-1}-\beta_f^{-1})e^{-2x^2r}}{[(\beta_i^{-1}-\beta_f^{-1})e^{-2x^2r}+\beta_f^{-1}]^2}.
\end{equation}
In the long-time limit the integral is well approximated by the one in which the upper extreme of integration is set to infinity,
and therefore \(\dot z^*\simeq t^{-d/2+1}\). Accordingly
\begin{equation}
z^*(s_c^{(eq,\beta_f)},t)\simeq z^*(s_c^{(eq,\beta_f)},\infty)+Ct^{-d/2},
\end{equation}
where the asymptotic value equals \(\beta_0\omega_0/2\) and \(C\) is a proportionality constant.
We conclude that
\begin{equation}
s_0(s_c^{(eq,\beta_f)},t)\simeq t^{d/2},
\end{equation}
namely Eq.~(\ref{divS0}).
We confirmed numerically the validity of this result in the specific case $d=3$.

\end{document}